\documentclass[fleqn,usenatbib]{mnras}

\usepackage[T1]{fontenc} 
\usepackage{ae,aecompl} 

\usepackage{graphicx} 
\usepackage{amsmath} 

\def\Mpc{\,{\rm Mpc}} 
\def\hmpc {\,h^{-1}\,{\rm Mpc}} 
\def\hunit {\, {\rm km \, s^{-1} \,Mpc^{-1} } } 
 
\def\zeq {z_{\rm eq}} 

\def\Omcb{\Omega_{\rm cb}} 
\def\Omm{\Omega_{\rm m}} 
\def\Omnu{\Omega_{\nu}} 
\def\Omlam{\Omega_{\Lambda}} 
\def\Omk{\Omega_{\rm k}}

\def\omc{\omega_{\rm c}} 
\def\omb{\omega_{\rm b}} 
\def\omcb{\omega_{\rm cb}} 
\def\omnu{\omega_{\nu}} 
\def\omm{\omega_{\rm m}} 
 
\def\omlam{\omega_{\Lambda}}

\def\ev {\, {\rm eV}} 
\def\xnu {x_{\nu}} 
\def\fnu {f_{\nu}}
\def\mnu {m_{\nu}} 
\def\sumnu {\Sigma \, m_{\nu}} 

\def\neff{ N_{\rm eff}}

\def\simlt{\la} 
\def\simgt{\ga} 
\def\newtwo { }  % bold-off. 

\title[ The CMB $m_\nu$ / vacuum energy degeneracy:
  deriving the slopes. ]
{ The CMB neutrino mass / vacuum energy degeneracy: 
   a simple derivation of the degeneracy slopes.}

\author[Will Sutherland] 
{Will Sutherland$^{1}$\thanks{E-mail: w.j.sutherland@qmul.ac.uk} 
\\
$^{1}$School of Physics and Astronomy, Queen Mary University of London, 
  Mile End Road, London E1 4NS 
}  % end of author-bracket. 

\begin{document}

\date{Submitted to MNRAS, 06 Sep 2017; revised 20 Feb 2018}

\pagerange{\pageref{firstpage}--\pageref{lastpage}} \pubyear{2018}

\maketitle

\label{firstpage}

\begin{abstract}
 It is well known that estimating cosmological parameters
 from cosmic microwave background (CMB) data alone results in a significant
  degeneracy between the total neutrino mass and several other 
  cosmological parameters,
  especially the Hubble constant $H_0$ and the 
  matter density parameter $\Omm$.   
  Adding low-redshift measurements such as baryon acoustic oscillations
 (BAOs) breaks this degeneracy and 
   greatly improves the constraints on neutrino mass. 
 The sensitivity is surprisingly high, e.g. adding the $\sim 1$ percent 
 measurement of the BAO ratio $r_s/D_V$ from the BOSS survey 
 leads to a limit $\sumnu < 0.19 \ev$, 
  equivalent to $\Omnu < 0.0045$   at 95\% confidence.  
  For the case of $\sumnu < 0.6 \ev$, 
  the CMB degeneracy with neutrino mass almost
  follows a track of constant sound horizon angle \citep{howlett12}.  
  For a $\Lambda$CDM + $\mnu$ model, we
  use simple but quite accurate analytic approximations to derive the slope
  of this track, giving dimensionless multipliers between the neutrino 
  to matter ratio ($\xnu \equiv \omnu / \omcb$) 
  and the shifts in other cosmological parameters. The resulting multipliers are
  substantially larger than 1: conserving the CMB sound horizon angle
  requires parameter shifts $\delta \ln H_0 \approx -2 \,\delta \xnu$,  
    $\delta \ln \Omm \approx +5 \, \delta \xnu$, 
   $\delta \ln \omlam \approx -6.2 \, \delta \xnu$, 
  and most notably $\delta \omlam \approx -14 \, \delta \omnu$. 
  These multipliers give an intuitive derivation of the degeneracy 
  direction, which agrees well with the numerical likelihood results 
  from {\em Planck} team.   

\end{abstract}


\section{Introduction}

  There is a long history of cosmological constraints on neutrino 
  masses; from the 1970s, 
  the simple requirement that the cosmic neutrino background 
  should not over-close the Universe required $\sumnu \simlt 50 \ev$ 
 \citep{cowsik72}.   
  This limit steadily improved with new data and simulations of
  large-scale structure during the 1990s,  with a notable improvement 
  to $1.8 \ev$ from the galaxy power spectrum in 
  the 2dFGRS survey \citep{elgaroy02}, and  
  the limit continued to improve through
   WMAP in 2003--2012 \citep{wmap9}. 
 
 Since the discovery of atmospheric neutrino oscillations by Super-Kamiokande
  \citep{superk} 
  showed that neutrinos have non-zero mass, 
 and the decisive solution of the solar neutrino problem by the Solar Neutrino
 Observatory \citep{sno02}, 
   many oscillation experiments
  with solar, nuclear reactor and accelerator neutrinos have given
 precise measurements of neutrino mass-squared differences \citep{rpp14}; 
 {\newtwo these imply
  $\sumnu \simgt 0.060 \ev$,   
  but do not set an absolute mass scale.}\footnote{ {\newtwo More formally, 
  the 95\% c.l. lower limits on neutrino mass-squared differences 
 imply a lower limit $\sumnu \ge 0.057 \ev$, 
  while the best-fit mass-squared differences 
  imply $\sumnu \ge 0.06 \ev$; these are so close that 
   $0.06 \ev$ is generally adopted as the fiducial minimum. }} 
  Current laboratory
 measurements give a model-independent upper limit $m_{\nu,e} < 2 \ev$
  for the electron neutrino \citep{rpp14},  
  while upper limits from cosmological observations 
  are now much stronger than this (though with some model-dependence). 

 Many previous works have studied the effects of neutrino mass on 
  CMB anisotropy, see e.g. \cite{mabert95}; 
 \cite{jkks96}; \cite{kks03}; \cite{bs04}; 
  \cite{lesg06}; \cite{hhrw11}; 
 \cite{howlett12}; \cite{hou13}; 
  \cite{riemer14}; 
 see also the reviews by \cite{lp06} and \cite{wong11} and the modern 
 textbook by \cite{lmmp13}, and references therein.   

 Currently, constraints from the {\em Planck} CMB data 
  alone \citep{planck15pars} provide 
 an upper limit $\sumnu < 0.6 \ev$ at $95\%$ confidence; this essentially
 requires that neutrinos remained quasi-relativistic until after the 
  epoch of CMB last scattering at $z_* \simeq 1090$, and in this 
  case the primary CMB anisotropies cannot improve much on this upper limit. 
   Secondary anisotropies, notably the gravitational lensing of the CMB, may 
  substantially improve the bound in the future 
 (\citealt{kks03}; \citealt{allison15}; \citealt{arch17}; 
  \citealt{core-sci} );     
  however in the {\em Planck} case adding lensing information only slightly
  changes the upper limit. {\newtwo (In more detail, adding lensing information
  does narrow the posterior, but also moves the likelihood peak from zero
  to positive neutrino mass;  the result is that the {\em Planck}-only 
  upper limit changes only slightly with the addition of lensing data).}  
  For future CMB experiments, the best sensitivity to neutrino mass 
  is anticipated from small-angle ($\ell > 1000$) polarization data,
  in a regime where the {\em Planck} data 
 is noise-limited \citep{planck15pars}.   
 
However, the CMB power spectrum alone gives a known degeneracy 
   between $\sumnu$ and low-redshift
 parameters such as $H_0, \Omm$, which can be broken by addition of
 low-redshift data such as baryon acoustic oscillations (BAOs) 
  sensitive to these parameters. 
   Combining {\em Planck}  with several BAO datasets 
  including BOSS \citep{boss}, 6dfGS \citep{sixdf} 
  and WiggleZ \citep{wigglez} 
  gives an upper limit $\sumnu < 0.17 \ev$ at 95\% confidence for a
 $\Lambda$CDM + $\mnu$ model (Eq. 54d of \citealt{planck15pars}), 
  or $0.23 \ev$ if polarization is not included (Eq.~57 of ibid).  
 The mid-point of these, $0.20 \ev$, is equivalent to a present-day 
 neutrino/matter ratio $\le 1.5\,$percent or $\Omnu < 0.005$ at 95\% c.l.,
  an impressively small limit given the $\approx 1$ percent
  precision of the most precise BAO measurement.  
   Even stronger limits 
  $\sumnu < 0.13 \ev$ have been derived using Ly-$\alpha$ forest data 
  (\citealt{palanque15}; \citealt{yeche17}), 
 but these are slightly more model-dependent.  

In this paper we give a simplified but fairly accurate semi-analytic
 derivation of the slope of this degeneracy track: in Section~\ref{sec-lever} 
  we note an interesting but not well-known feature 
   that the CMB sound horizon angle is
  approximately $2.5\times$ {\em more} sensitive to small changes in neutrino 
 density compared to CDM+baryon density;  we then 
  estimate various dimensionless multipliers relating parameter 
 variations along the CMB-only degeneracy track. 
  In Section~\ref{sec-num} 
 we compare these analytic approximations with 
   numerical results, including the public {\em Planck}
 likelihood results. In Section~\ref{sec-pk} we
 consider effects on the matter power spectrum, and note that
  the secondary effects from varying $H_0$ turn out of similar size
   to the primary effects of neutrino mass.  
   In Section~\ref{sec-8par} we briefly discuss
  extended models, and we conclude in Section~\ref{sec-conc}.  
  
\section{The lever-arm between neutrino mass and low-redshift
  parameters} 
\label{sec-lever} 

In this section we give a simple derivation of the lever-arm between 
 the present-day neutrino/matter density ratio to
 low-redshift parameters such as $\Omm, \Omlam, H_0, \omlam$, defined below.   

\subsection{Notation} 
 Our default model is flat $\Lambda$CDM extended with 
 arbitrary neutrino mass, unless specified otherwise. 
 We use the standard notation that $h \equiv H_0 / (100 \hunit)$,
 and $\Omega_i$ is the present-day density of species $i$ in units 
 of the critical density, where $i = {\rm c,b,\nu,\Lambda}$ respectively 
  for CDM, baryons, neutrinos and the cosmological constant.  
 The physical densities $\omega_i$ are defined by 
  $\omega_i \equiv \Omega_i \, h^2$. We assume zero curvature 
 $\Omega_k = 0$, dark energy equation of state $w = -1$,
   and effective number of neutrino species $\neff = 3.046$, 
  except in Section~\ref{sec-8par} where we briefly explore deviations
  from these. 

  We use $\Omcb \equiv \Omega_c + \Omega_b$
 to denote the dark + baryonic matter density (excluding neutrinos), 
  $\Omm \equiv \Omcb + \Omnu$ includes neutrinos, 
 $D_* \equiv (1+z_*) D_A(z_*) \approx 13.9\, {\rm Gpc}$ 
  is the comoving angular diameter distance to
   photon decoupling at redshift $z_* \simeq 1090$, 
  $\theta_* \equiv r_S(z_*) / D_*$ is the CMB sound horizon angle, 
  and $z_{eq} \approx 3375$ is the redshift of matter-radiation
  equality.  

 It is helpful below to work mostly with dimensionless parameters, 
  so we define the present-day neutrino / other matter ratio as 
\begin{equation} 
\label{eq:xnu} 
  \xnu \equiv \omnu / \omcb \ \ ; 
 \end{equation} 
  note that a more common parameter 
  choice is $f_\nu \equiv \omnu / (\omcb + \omnu) 
   = \xnu / (1 + \xnu)$ where $f_\nu$ includes neutrinos in the
  denominator; these are clearly 
  very similar for $\xnu, f_\nu \ll 1$, but it is 
  convenient later to choose a parameter which is strictly linear 
   in $\sumnu$ for fixed $\omcb$.  
 For the concordance value $\omcb \simeq 0.141$, this gives 
  $\xnu = \sumnu / (13.1 \ev)$, and a default value 
  (for $\sumnu = 0.06 \ev$) of $\xnu \simeq 4.6 \times 10^{-3}$. 
 Since we are mostly interested in {\em differences} in observables 
  relative to the 6-parameter model with neutrino masses fixed to
 the default,   we also define $\delta\xnu \equiv  
  \xnu - 0.0046$ to be the shift in $\xnu$ above this 
 minimal value. 

\subsection{Neutrino effects on the sound horizon length} 
\label{sec:soundhor} 

 If the total neutrino mass is $\sumnu \simlt 0.6 \ev$
  (a conservative limit from {\em Planck} data alone),   
 then the oscillation experiments require all three single neutrino masses
 $\le 0.22 \ev$. At high redshift the neutrino temperature 
  is $T_\nu \simeq (4/11)^{1/3} T_\gamma$ where $T_\gamma$ is the photon
 temperature, hence at photon decoupling we have $T_\nu(z_*) = 2122 \, {\rm K}$ 
 and $k T_\nu(z_*) = 0.183 \ev$.  From the accurate fitting functions in
  Sect.~3.3 of \cite{wmap7}, 
 each single neutrino with $m_\nu = 0.183 \ev$ 
 would contribute 6.5 percent higher energy density at decoupling 
 than one negligible-mass neutrino, which is a quite substantial shift.
 {\newtwo However, the effects of neutrino mass on the sound horizon 
   length are suppressed by several factors as follows:} 
   since minimal-mass neutrinos contribute 10.0~percent of the total 
  matter+radiation density at
 $z_* \simeq 1090$, changing to 
  $\sumnu = 0.55 \ev$ ($\delta\xnu = 0.037$, i.e. three neutrinos
   with masses close to $0.183 \ev$ each) 
   gives only a 0.65~percent increase in 
  total energy density at $z_*$, thus 0.32~percent increase in 
  expansion rate $H(z_*)$.  
  Finally, the sound horizon length $r_S(z_*)$ contains an integral over 
  $\infty > z > z_*$, and the fractional shift in $H(z)$ decreases
   towards higher redshift, so the change in sound horizon length  
  is smaller again at $-0.15$~percent. Also for
  $\sumnu < 0.55 \ev$ the fractional effect falls faster than linearly, 
  becoming almost negligible at $\sumnu \simlt 0.3 \ev$.  
  (See also Section~\ref{sec-num} for a numerical verification of
    the above). 

 However, neutrino mass does have important effects on $D_*$ and hence 
  $\theta_*$ as we see below.  

\subsection{Do massive neutrinos affect matter-radiation equality ?} 

 The short answer is ``very little'', in the case of 
  fitting the {\em Planck} data.
 Traditionally, many early works studied the consequences of
  varying $\sumnu$ at fixed $\Omm, h$, in which case the CDM density
  is implicitly reduced 1:1 as neutrino density increases; however this  
  affects the CMB by altering the epoch of matter-radiation equality, $\zeq$ 
 and also shifts $\theta_*$ as we see below; so the observable degeneracy
  track is substantially different to fixing $\Omm, h$.  

 For neutrino masses below $\sumnu < 0.6 \ev$ or single neutrino masses
 $< 0.2 \ev$, the shift in neutrino energy density (relative
  to minimal-mass neutrinos) around $\zeq \sim 3375$ 
 is no more than 1 percent, and the shift in radiation (photon + neutrino)
 density is $\approx 0.41 \times$ this hence $\le 0.41$~percent, which is 
  substantially smaller than the Planck precision on $\zeq$.  
 Thus the ``direct'' effect of neutrino mass around 
  $z \sim 3000$ has very little impact on 
  $\zeq$, and any change in $\zeq$ is driven mainly by 
  any consequential shift in $\omcb$, which turns out to be small
 in the {\em Planck} case (see \S~\ref{sec-num}).  

 If we adopt the common choice of $\omm$ and $f_\nu$
 among the base cosmological parameters, clearly $\omcb \equiv \omm (1-f_\nu)$, 
 so raising $f_\nu$ at constant $\omm$ trades CDM for
 neutrino density today in equal ratio;   
 in that case increasing $f_\nu$ clearly does reduce $\zeq$ 
  nearly in proportion.  
  This has the apparent benefit of keeping $D_*$ (and {\newtwo also 
 the age of the universe}, $t_0$ ) 
  almost constant as $\fnu$ varies, but this benefit is largely
   illusory, since the change in $\omcb$ also changes the sound horizon length 
  and hence $\theta_*$ (see below); and it is $\theta_*$ which is 
  constrained most precisely by {\em Planck} data, 
   rather than $D_*$ or $t_0$.   
 
 Thus we argue that $\omm$ and $\fnu$ are not an optimal choice
  for basic parameters, and a more natural choice is to 
  use $\omcb$ and $\xnu$; 
  so $\omm \equiv \omcb (1+\xnu)$ becomes a derived
 parameter. This is preferable since varying $\xnu$ up to 
  $\simlt 0.04$ at constant $\omcb$ has a nearly negligible  
   effect on $\zeq$.  
  
 In any of these parameter choices the sound horizon angle 
  $\theta_*$ does vary with $\fnu$ or $\xnu$: it turns out that this can
 only be compensated by a change in vacuum energy density and hence $h$, 
  for reasons given below.  

\subsection{Sensitivity of $\theta_*$ to neutrino and CDM density} 

 Simple intuition suggests that increasing neutrino mass 
  should be compensated by a reduction in CDM density to conserve
 consistency with CMB data. This intuition turns out to be incorrect, 
  for the following reasons. 

 The observed sound horizon angle $\theta_* \equiv r_S(z_*) / D_*$ 
   (where $r_S(z_*)$ is the comoving sound horizon length at last scattering) 
  is the most precise cosmological observable (apart from the absolute
  temperature $T_{0}$):
   $\theta_*$ is constrained to 0.06 percent precision by {\em Planck}
 \citep{planck15pars}, 
  and the corresponding length $r_S(z_*)$ is also well constrained
   at 0.25 percent precision, since the latter follows from the measurements
    of $\omc$ and $\omb$ from the acoustic peaks. 
 (Given the high precision on $\omb$ from {\em Planck}, 
    variations in $\omb$ have very little effect on $r_S(z_*)$, 
  so in practice it is the combined value $\omcb$ which is 
   relevant below).  
 Thus,  if we vary $\xnu$, to remain consistent with 
 the {\em Planck} data it is necessary to 
  vary other parameter(s) to preserve a near-constant angle $\theta_*$.  
  This degeneracy is studied in detail numerically by \cite{howlett12}, 
 and is found to be well represented by constant $\omcb$ and $D_*$ as above. 
 As seen above, for the interesting range 
  $0.06 < \sumnu < 0.6 \ev$ ($\xnu \simlt 0.046$),  
  varying the neutrino mass has nearly negligible
   effect on the sound horizon length and the heights
 of acoustic peaks; but it does have a significant 
  effect on the distance $D_*$ 
  to last scattering, since the heaviest neutrino(s) must have
  mass $> 0.05 \ev$ and became non-relativistic at $z \simgt 250$, thus 
  increasing the expansion rate during most of the post-recombination era. 

 To get a semi-analytic estimate of this degeneracy track, 
  a good approximation to the present-day horizon size
  in flat-$\Lambda$ models was given by \citet{vs85} as 
 \begin{equation}
  r_H \simeq \frac{2 \, c}{H_0 \, \Omm^{0.4}} \ . 
 \end{equation} 
 The value of $D_*$ is about 1.8 percent smaller than the above
  due to the finite redshift of last scattering, which 
  leads to 
 \begin{eqnarray} 
  D_* &\simeq & \frac{ 5888 \Mpc } {h \, \Omm^{\,0.4}} 
 \label{eq:dstar1} 
\end{eqnarray} 
 this is accurate to $< 0.1$ percent
  for the range of $\Lambda$CDM + $\mnu$  
 models allowed by {\em Planck}.  This small error is fairly 
  unimportant in the following, since it is smaller than the 
  $\sim 0.25$ percent observational uncertainty  in $r_S(z_*)$ and hence $D_*$.
 It is convenient to rewrite this as 
\begin{eqnarray}  
%      & \simeq &  \frac{ 5888 \Mpc} { \omm^{0.4} \, h^{0.2} } 
 D_* &\simeq & \frac{ 5888 \Mpc } 
   { \omcb^{\,0.4} \, (1 + \xnu)^{0.4} \, h^{0.2} } \ , 
\label{eq:dstar2} \\  
  &\simeq & \frac{ 5888 \Mpc } { \omcb^{\,0.4} \, (1 + \xnu)^{0.4} \, 
  (\omcb + \omnu + \omlam)^{0.1} } \ ; 
\label{eq:dstar3} 
\end{eqnarray} 
% \footnote{The main
% source of error is that neutrinos are not quite matter-like 
%  over the interval $300 < z < 1000$.  In this interval, the massive
% neutrinos contribute slightly {\em more} energy density compared
% to either today's equivalent
% amount of CDM {\em or} a massless neutrino, but slighly {\em less}
% than the sum of these. }   

 Thus, if $\xnu$ increases from its minimal value $\xnu \simeq 4.6 \times
  10^{-3}$, we must adjust other parameter(s) to restore $\theta_*$ to the
  precisely-measured {\em Planck} value. 
 At first sight it appears we could reduce $\omcb$ to compensate, 
  but we now illustrate qualitatively that this does
  not lead to an acceptable solution.  
 Concerning variations in $\omcb$, 
  although the distance $D_*$ does scale 
   as $\omcb^{-0.4}$ (for fixed $h$), varying
  $\omcb$ also produces a shift in the sound horizon length 
   as $r_S(z_*) \propto \omcb^{-0.25}$ which partly compensates, so the 
  net sensitivity of $\theta_*$ to $\omcb$ becomes  
 \begin{equation} 
\label{eq:theta-omcb} 
  \left( \frac {\partial \ln \theta_*}{\partial \ln \omcb} \right)_{\xnu,h} 
  \simeq  +0.15 \ ;
 \end{equation} 
 where the subscripted parameters are held fixed.    
%  see e.g. \citet{psp02} for more details.\footnote{ {\newtwo Note that 
% \citet{psp02} assumed zero neutrino mass so their $\omm$ is equivalent
%  to our $\omcb$.  As above, neutrino 
%  mass has very weak effect on the sound horizon length, so it is essentially
%   $\omcb$ which is relevant to the sound horizon length but $\omm$ 
%  which determines $D_*$. }}  

 However, varying neutrino mass has (almost) negligible compensation 
  from $r_S(z_*)$; small
  neutrino masses ($\sumnu < 0.6 \ev$) affect $D_* \propto (1+\xnu)^{-0.4}$ 
   but have nearly negligible effect on sound horizon length, hence 
 \begin{equation} 
\label{eq:theta-omnu} 
  \left( \frac {\partial \ln \theta_*}{\partial \xnu} \right)_{\omcb,h} 
   \simeq  +0.4 \ .
 \end{equation} 
 (Numerical differentiation with CAMB actually gives +0.34 rather
   than 0.40, see Section~\ref{sec-num} below for more details). 
 Note that if we fix $\omm$ and vary $f_\nu$,
  then we get the difference of these,  i.e. 
 \begin{equation} 
  \left( \frac {\partial \ln \theta_*}{\partial \fnu} \right)_{\omm,h}  
   \simeq  +0.25 \ .  
 \end{equation} 
From Eq.~\ref{eq:dstar2} we also have the sensitivity to $h$ as 
\begin{equation} 
  \left( \frac {\partial \ln \theta_*}{\partial \ln h} \right)_{\omcb,\xnu}  
   \simeq  +0.2 \ .  
\end{equation}

% There are several interesting points to note from the above:
% \begin{enumerate} 
% \item The sound horizon {\em length} is nearly insensitive to neutrino mass; 
% \item  Neutrinos and CDM density have nearly identical effects on $D_*$, as
% expected;    
% \item However, the effects of CDM density on $D_*$ and $r_S(z_*)$ partially 
%  compensate each other in $\theta_*$, 
%  while the similar neutrino effect on $D_*$ has negligible 
%  compensation from $r_S(z_*)$. 
% \end{enumerate} 
 Although all of the $\theta_*$ sensitivity coefficients above are 
  fairly small compared with 1, 
  the {\em Planck} estimate of $\theta_*$ is much more
 precise than any other parameter, 
  so it is the {\em relative} sizes of these coefficients
   which mainly determine the direction of the CMB degeneracy track. 
 A notable point above, comparing (\ref{eq:theta-omcb})
 and (\ref{eq:theta-omnu}), 
  is that $\theta_*$  is more than twice as sensitive 
  to a small change in neutrino density than an equal shift 
   in CDM+baryon density;  the effects on $D_*$ are similar, but the
  CDM effect on $\theta_*$ 
  is substantially compensated by variation in sound horizon length, 
   while the effect from $\xnu$ is almost uncompensated.  

  This turns out to be a major reason (see below) 
  why increasing neutrino mass {\em cannot} in practice be 
   compensated by reducing dark matter density $\omega_c$, 
 but instead requires a (considerably larger) reduction in 
  dark energy density.  

 If we take an example case of $\delta\xnu = +0.01$ 
 (i.e. increasing $\sumnu$ from $0.06\ev$ to $0.191 \ev$, 
  thus near the current CMB+BAO upper limit) 
  we can consider three illustrative cases for varying $\omcb$: 
 \begin{enumerate} 
 \item If $\omcb$ and $h$ were both held fixed 
  then the above shows that $\theta_*$ would increase by $\approx 0.4$ percent, 
   which is over $6\times$ outside the {\em Planck} precision.  
 \item We may compensate the change in $\xnu$ with an equal 
   (1 percent) reduction in
   physical matter density $\omcb$, thereby keeping constant $\omm$, $h$ 
   and almost constant $D_*$;  
  such a shift in $\omcb$ and $\zeq$ would be tolerable at
   around the 1$\,\sigma$ {\em Planck} precision. 
   However, due to the differing sensitivities above, 
   this would give a $\sim +0.25$~percent increase in $\theta_*$, 
   which is still $\sim 4\times$ larger than the {\em Planck} precision
   and therefore ruled out.  
 \item Finally, we could reduce $\omcb$ by a larger percentage in order 
    to conserve $\theta_*$ at its fiducial value.  
  From above, this would require 
   $\sim 2.5$ percent reduction in $\omcb$ and $\zeq$;  
  however, a shift this large in $\zeq$ would lead to 
   substantial tension with the acoustic peak heights. 
\end{enumerate} 
 
 The above example shows that if 
  $\delta\xnu \simgt +0.01$ (and $\omlam$ or $h$ were held fixed),  
  an arbitrary adjustment to $\omcb$
  could conserve either $\theta_*$ or $\zeq$ within the {\em Planck} bounds,
  but {\em not} both simultaneously.    
  Thus, the only way to preserve consistency with the
  {\em Planck} data is to conserve $\omcb$ and $\zeq$, but 
  to reduce the physical dark energy density $\omlam$ 
 (thereby reducing $h$) to preserve the concordance value of $D_*$;
  we estimate the resulting degeneracy direction in the next subsection.  

% FFF Figure here. 
\begin{figure*} % [ht] 
\begin{center} 
\includegraphics[angle=-90,width=15cm]{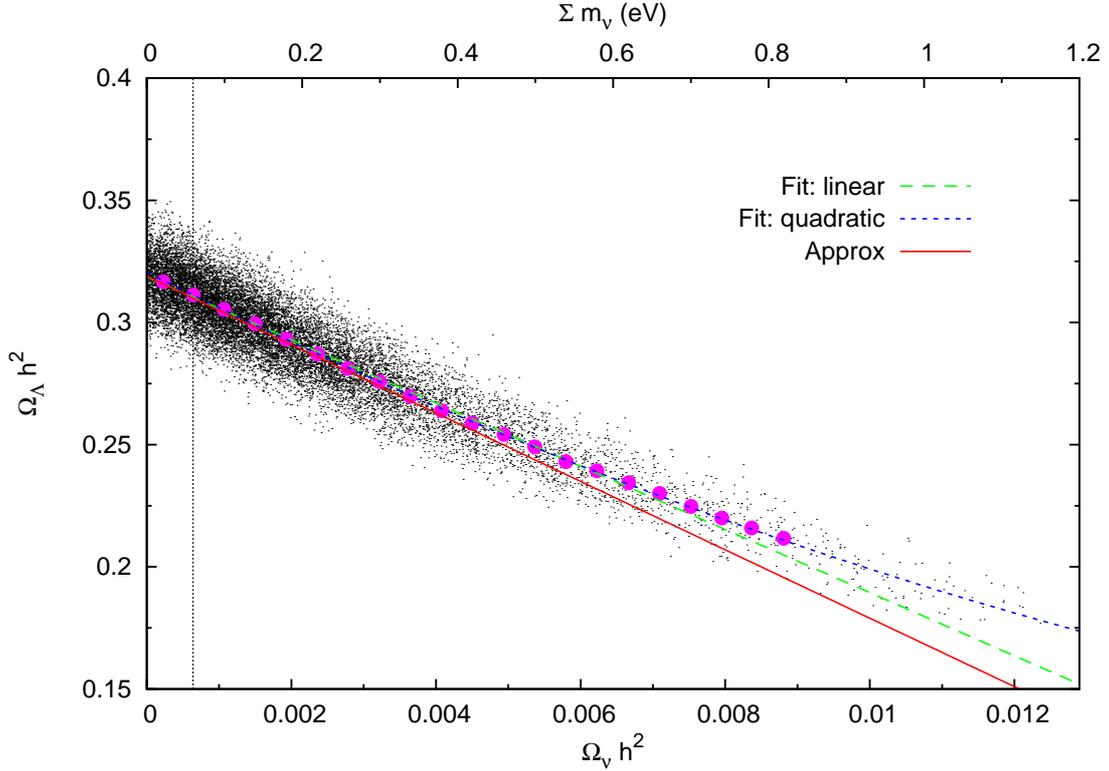} 
\caption{
The correlation of physical vacuum energy density 
  $\omlam$ versus neutrino density $\omnu$ for CMB-compatible models.  
 Small dots show a random subsample of 
  models from a {\em Planck} MCMC chain; large points
  show the mean values, marginalised in $0.04 \ev$ wide bins of $\sumnu$.  
 The solid line shows the slope $-14$ from approximation~\ref{eq:wlam-wnu},
 the long-dashed line is a linear fit to the Planck points, and 
 the short-dashed line is a quadratic fit to the Planck points. 
  The vertical dotted line is the minimal value $\sumnu = 0.06\ev$. 
\label{fig-wlam} 
} 
\end{center}
\end{figure*}  

\subsection{The lever-arm from neutrino mass to dark energy} 

 If $\omcb$ is fixed by the CMB acoustic peak heights 
  and a value for $\xnu$ is assumed,
  then {\newtwo (for a flat-$\Lambda$ model)}
  specifying any one of $\omlam, h, \Omcb, \Omlam$ determines 
  the other three: so any of those four may be adopted as
  the independent ``low-redshift'' variable, and the degeneracy
 track is approximately a line through a five-dimensional 
  $\xnu, h, \omlam, \Omcb, \Omlam$ space.  We now estimate 
 the direction of this track. 

 It is easily seen from Eqs.~\ref{eq:dstar2} and~\ref{eq:dstar3}
  that a small increase $\delta \xnu \equiv \xnu - 0.0046$ 
  above the minimal value requires relative changes
 \begin{eqnarray} 
\label{eq:wtot} 
 \delta \ln (\omcb + \omnu + \omlam) &\simeq & -4 \, \delta \xnu 
 \qquad { \rm and} \\  
\label{eq:hxnu} 
 \delta \ln h &\simeq & -2 \, \delta \xnu 
 \end{eqnarray}  
  to conserve $D_*$ at its fiducial value. 

 From these the consequential shifts in
  other parameters readily follow as 
 \begin{eqnarray} 
  \delta \ln \Omcb \simeq +4 \, \delta \xnu \ ; \qquad  
  \delta \Omcb \simeq +4 \, \delta \Omnu \ ; \\ 
 \delta \Omm \simeq +5 \, 
  \delta \Omnu\ ; \qquad \delta \Omlam \simeq -5 \, \delta \Omnu
 \end{eqnarray} 
 here 
  we have chosen ``matching units'' i.e. $\ln(\Omcb)$  
  vs dimensionless neutrino fraction $\xnu$,  
  or $\Omega_i$ on both sides.  

 It is notable that  
  the dimensionless multiplier 
  is dramatically larger in physical density 
 units,  $\omlam$ vs $\omnu$: 
  rearranging Equation~\ref{eq:wtot} we have 
\begin{eqnarray} 
%  \delta \ln (\omcb + \omnu + \omlam) & \simeq & -4 \, \delta \xnu  \\ 
 \delta (\omcb + \omnu + \omlam) &\simeq & 
     -4 \, \frac{ \delta \omnu} { \omcb}  \, ( \omcb + \omnu + \omlam) 
 \nonumber \\ 
  &\simeq & \frac{-4}{\Omcb} \, \delta \omnu \nonumber \\ 
\delta \omlam &\simeq & \! \left( \frac{-4}{\Omcb} - 1 \right) \, \delta \omnu 
  \simeq  -14  \, \delta \omnu 
 \label{eq:wlam-wnu} 
\end{eqnarray} 
  where the last line assumes $\delta \omcb \simeq 0$.  

 For the relative change in $\omlam$, we can 
   divide the above by $\omlam$ and rearrange to  
\begin{equation} 
 \delta \ln \omlam \simeq \left( \frac{-4 - \Omcb}{\Omlam} \right) \delta \xnu 
 \quad  \simeq -6.2 \, \delta \xnu  \ . 
\end{equation} 

The various dimensionless multipliers above are notably larger than 1, 
 with many between 4 to 6, and the surprisingly large factor $-14$ in 
 physical densities $\delta\omlam \simeq -14 \, \delta\omnu$. 
This is arguably the ``root cause'' of the degeneracy in physical
 density units,  i.e. increasing neutrino 
  density requires a $-14\times$ larger reduction in vacuum energy
  density to minimise the changes in the observed CMB power spectrum. 

 Qualitatively, this is explained because an increase in neutrino
 mass increases the expansion rate across almost the entire 
  matter-dominated era (causing a decrease in $D_*$); 
 to restore $D_*$ we must then reduce 
 the expansion rate $H(z)$ in the $\Lambda$-dominated (accelerating) era 
 $z \simlt 0.67$ by reducing $\omlam$. 
  The latter era contributes nearly half of cosmic time, but only about
  19 percent of the comoving distance $D_*$ to last scattering, 
   roughly explaining the ``factor of 4'' in Equation~(\ref{eq:wtot}).
  Another factor
   of $\Omcb^{-1}$ appears in Equation~(\ref{eq:wlam-wnu}),
  since we defined $\xnu$ relative to the CDM+baryon density while 
  $\delta \ln (\omcb + \omnu + \omlam)$ is 
   relative to the total mass-energy density today.  

 Although the {\em physical} CDM+baryon density $\omcb$ changes
 very little in reponse to varying neutrino mass, the
   density parameter $\Omm$ is affected substantially:  
  assuming flatness we can write 
 \begin{equation} 
 \Omm = \frac{\omcb + \omnu}{\omcb + \omnu + \omlam} \ ,  
\end{equation} 
  so it is clear that as $\omnu$ increases,  
  it is the steep decrease in dark energy density 
 in the denominator which is
 reponsible for around four-fifths of the increase in $\Omm$, 
  while the ``direct'' contribution of $\omnu$ 
  in the numerator accounts for only the remaining one-fifth. 

 In the next subsection we compare these approximate estimates with
  selected numerical results from the {\em Planck} collaboration 
 likelihood chains and the literature, and find rather good agreement.

\section{Comparison with numerical results} 
\label{sec-num} 

The derivations above are approximate, but turn out to
 be quite close to the numerical degeneracy direction 
  in current and near-future CMB experiments, as follows.  

 Numerical calculations with CAMB indicate that 
  if $\omc, \omb$ are held fixed while $\sumnu$ is varied,
   then the condition for constant $\theta_*$ is actually 
  $\delta \ln h \simeq -1.75 \, \delta \xnu$; 
  this slope is comparable but slightly shallower 
  than the value $-2$ from approximation~\ref{eq:hxnu} above. 
  The difference between $-2$ and $-1.75$ arises mainly from two small
  effects: 
   neutrinos are not fully matter-like at $300 \la z < 1090$ which 
   makes the sensitivity of $D_*$ to $\xnu$  
  slightly weaker than the $-0.4$ power in Eq.~\ref{eq:dstar3} above, 
   actually $\simeq -0.36$; also increasing   
  neutrino mass very slightly reduces the sound horizon 
   length $r_S(z_*)$. (The main point remains, that
   $\theta_*$ is substantially more sensitive to $\xnu$ than
   $\omcb$ and $h$).  

 However, in full 7-parameter CMB fits, varying $\xnu$ also gives 
  additional small correlated changes in other parameters, 
  with $\omcb$ being the next most important contributor to 
   changes in $\theta_*$: changes in $\omb$ are much less important.  
  In the full parameter space, 
  the condition for constant $\theta_*$ is well approximated by 
  the relationship 
\begin{equation} 
\label{eq-hxnuwcb} 
    \delta \ln h \simeq -1.75 \, \delta \xnu - 0.8 \,\delta \ln \omcb \ .  
\end{equation} 
 In the {\em Planck} case,  the likelihood ridge shows a small but positive
  correlation\footnote{ {\newtwo The source of this correlation 
   appears to be that when increasing neutrino mass at 
   constant $\omcb, \theta_*$, the largest fractional change
   of the theoretical CMB spectrum is a small reduction at low multipoles, 
    since the higher $\Omm$ and thus lower $\Omlam$ 
  reduces the late-time ISW effect. 
  The shift is well within the cosmic variance, but in the MCMC fits 
   this can be partially compensated by
  a combination of small red tilt (lowering $n_s$) and a small 
   increase in $\omcb$.} } 
 of $\omcb$ with $\xnu$,  
   in the direction $\delta \ln \omcb \approx +0.2 \, \delta \xnu$:  
  so marginalising over $\omcb$ leads to an overall degeneracy direction  
 for {\em Planck} of $\delta \ln h \approx -1.9 \, \delta \xnu$, 
  hence fortuitously moving closer to the simple $-2$ approximation 
   from the previous Section.    
%%qqq done to here. 
 
  Fig.~\ref{fig-wlam} shows a scatter plot of  
  $\omlam$ vs $\omnu$ in the public Planck
 Monte-Carlo Markov chains, here the 7-parameter chain 
  {\tt base\_mnu/plikHM\_TTTEEE\_lowTEB }. Fitting a linear
 relation for $\omlam$ vs $\omnu$ to the Planck chain gives a
 slope of $-12.9$, while a quadratic fit has a slope of $-14.4$
  at the fiducial $\sumnu = 0.06 \ev$; the latter in particular
 agrees well with the approximate slope $-14$ from Eq.~\ref{eq:wlam-wnu}. 
 Here the quadratic is a better fit, since the observed degeneracy 
  track starts to curve for $\sumnu \simgt 0.4 \ev$ 
 where neutrinos are not fully relativistic at recombination; 
  this pulls the linear fit to shallower slope.   

 Looking to the future, a detailed set of predictions 
   for the proposed {\em CoRE} CMB spacecraft \citep{core-sat} 
 are given by \cite{arch17} and \cite{core-sci}.  
 In the case of simulated {\em CoRE} data, the improved sensitivity 
  to high-$\ell$ polarisation and CMB lensing 
  leads to a substantially more positive 
  correlation of $\omcb$ versus $\xnu$ for
 {\em CoRE} than for {\em Planck}, {\newtwo 
  with a predicted {\em CoRE} likelihood ridge given by }   
  $\delta \ln \omcb \approx +1 \, \delta \xnu$.    
  Also, Eq.~2.4 of \cite{arch17} converts to 
\begin{equation} 
  \delta \ln h \simeq -2.5 \, \delta \xnu  
\end{equation}  
 in our notation, {\newtwo which is consistent with 
 substituting $\delta \ln \omcb \approx +1 \delta \xnu$ into
  approximation~(\ref{eq-hxnuwcb}) 
 above}.   This predicted degeneracy slope -2.5 for {\em CoRE} is 
  somewhat steeper than the {\em Planck} case and our 
  approximate slope -2 above. However,  
  it is not dramatically steeper, because the larger 
  coefficient of $\delta \xnu$ in approximation~(\ref{eq-hxnuwcb}) 
  {\newtwo implies that the $\delta \xnu$ term still 
    dominates over the $\omcb$ term}.     

We can also compare with the numerical estimates of \cite{pan-knox},
 who show $H(z)$ for CMB-fitted models with several 
 selected values of $\sumnu = (0.05, 0.1, 0.2) \ev$ imposed as a prior.
  At $z \simlt 100$ we can write  
\begin{equation} 
  H(z) = 100 \hunit \sqrt{ (\omcb + \omnu)(1+z)^3 + \omlam} \ . 
\end{equation} 
From the condition in Equation~(\ref{eq:wlam-wnu}),  
   $\delta \omlam \simeq -14 \,\delta \omnu$, 
 it is clear that this predicts  
  a `crossover' redshift given by $1 + z_{cr} \simeq \sqrt[3]{14}$ 
  i.e. $z_{cr} \simeq 1.4$, at which the CMB-preferred value of $H(z_{cr})$ 
 becomes independent of neutrino mass. 
  The numerically-fitted  
   crossover point seen in Figure~1 of \cite{pan-knox} agrees very 
  well with this simple estimate. 

{\newtwo To verify that the direct dependence of the sound horizon length 
  on neutrino mass is nearly
 negligible, we did a fit of $r_S(z_*) (\omcb/0.1410)^{0.25}$ as a quadratic
 function of $\sumnu$ using the above Planck chain.  
  The scaling with $\omcb$ is included in order to 
  cancel the secondary effect from correlated shifts of $\omcb$ with 
  $\sumnu$, which otherwise dominate the variation in $r_S(z_*)$ alone.  
 The result of this fit is 
\begin{eqnarray} 
\label{eq:rsmnu} 
%% The lefteqn thing shifts alignment 
\lefteqn{ 
 r_S(z_*) \left( \frac{\omcb}{0.1410} \right)^{0.25} / (1 \,\rm {Mpc}) 
 }
  \nonumber \\
   & = & 144.84 + 0.078 \left( \frac{\sumnu}{1 \ev} - 0.06 \right) \nonumber \\
   & & - 0.511 \left( \frac{\sumnu}{1 \ev} - 0.06 \right)^2 
\end{eqnarray} 
  with the rms of 
  the Planck chain only 0.076 Mpc (0.05 percent) relative to the above fitting
  function. The fit gives a mean shift of $< 0.01$ percent 
    for $\delta\xnu = 0.01$,
  and $-0.08$~percent for $\sumnu = 0.6\ev$ ($\delta\xnu = 0.04$).  
 This validates the argument in Section~\ref{sec:soundhor} that
  neutrino mass has nearly negligible direct 
  effect on sound horizon {\em length}. 
} % end new bit. 
 
 To summarise this section, we expect that numerical degeneracy tracks 
  from CMB experiments alone will in general give a track with slope  
   $\delta \ln h / \delta \xnu$ between $-1.75$ and $-2.5$ in order to 
  conserve the sound horizon angle $\theta_*$.    
   Here the simple slope estimate $-2$ from 
   approximation~(\ref{eq:wtot}) is the leading-order term,
  while smaller effects from errors in that approximation 
  and correlations between $\omcb$ and $\xnu$ give moderate corrections
   to the simple $-2$.  In physical density units, the slope of 
 $\delta \omlam$ vs $\delta \omnu$ is approximately $7\times$ steeper 
  than the above.  

\section{Consequences for the matter power spectrum} 
\label{sec-pk} 

% FFF Figure here. 
\begin{figure} % [ht] 
\begin{center} 
\includegraphics[angle=-90,width=9cm]{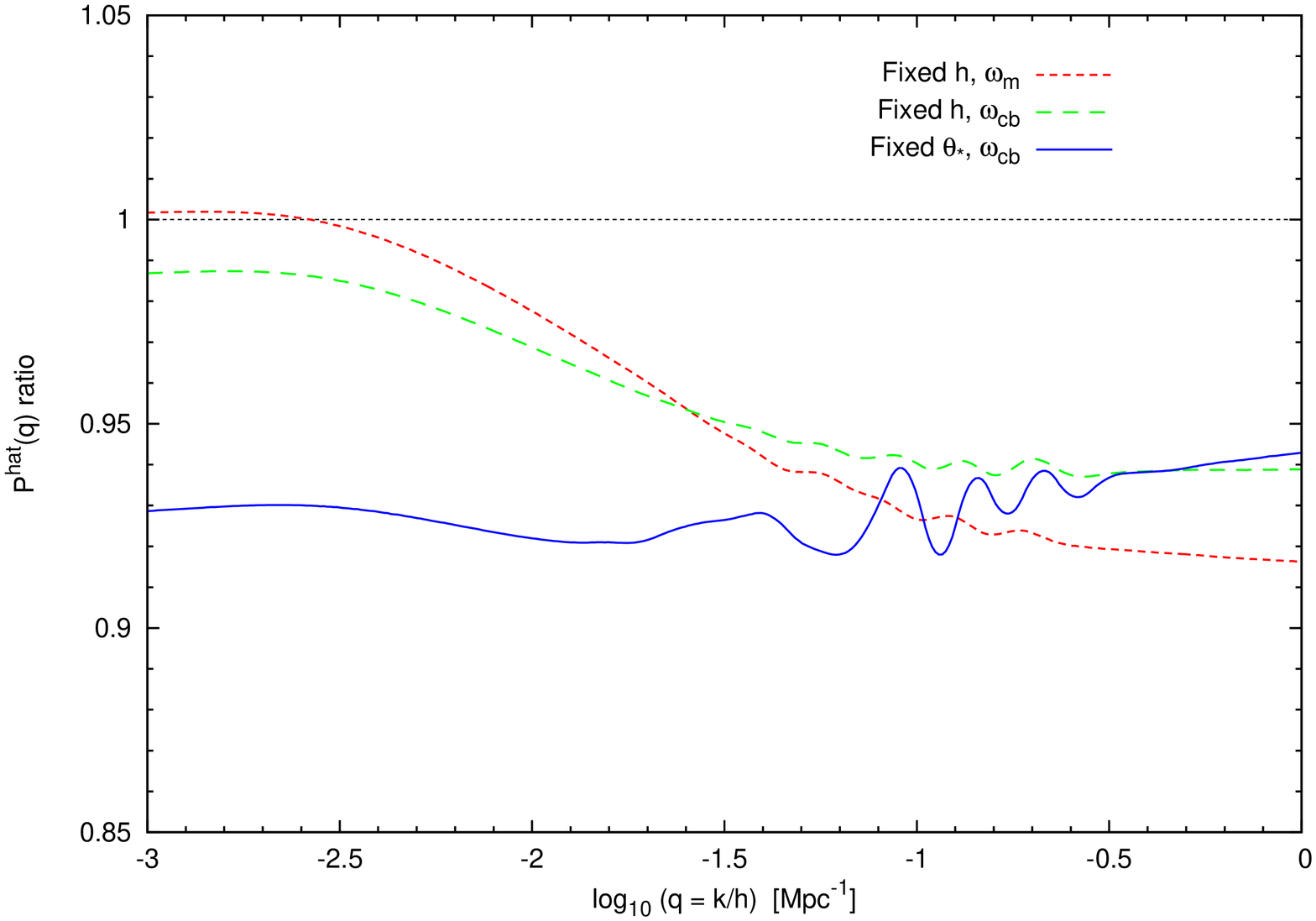} 
\caption{
 Linear-theory matter power spectrum for three models, all
  with $\delta \xnu = +0.01$ ($\sumnu = 0.191 \ev$), relative 
  to the baseline model with $\delta \xnu = 0$ ($\sumnu = 0.06 \ev$). 
   The lines show three
  choices for which other parameters are fixed: the short-dashed line has 
 fixed $\omm, h$ (i.e. $\delta \omcb = -1 \, \delta \omnu$); 
 the long-dashed line has fixed $\omcb, h$ (i.e. 
  $\delta \omlam = -1 \, \delta \omnu$); and the solid line has fixed
 $\omcb, \theta_*$ (i.e. $\omlam$ and $h$ reduced to conserve $\theta_*$)  
   approximating the {\em Planck} degeneracy track.  
\label{fig-matpow} 
} 
\end{center}
\end{figure}  

The degeneracy between $\sumnu$ and $h$ turns out to have
 interesting consequences for matter power spectrum observables,
  as follows.  
There is a well-known effect that massive neutrinos 
  reduce the matter power spectrum on small scales (large wavenumber $k$) 
 due to neutrino free-streaming; 
 there is negligible effect at $k < k_{fs}$ where $k_{fs}$ is the
  free-streaming scale, while the suppression shows
 a downward ramp at $k > k_{fs}$, then asymptotically
 approaching $\delta P / P \approx -8 \,f_\nu$ on small scales 
  \citep{het98},
 as in the short-dashed line in Figure~\ref{fig-matpow}.  
  However, most early studies (e.g. \cite{eh97}; \cite{lp06})
   compared models with different $\sumnu$ but identical $\Omega_m, h$
   to derive this simple rule-of-thumb:   
 if instead we compare models with different $\sumnu$ moving
  along the CMB degeneracy track, 
 then the resulting variations in $h$ and $\Omm$ also become comparably 
  important {\newtwo for the low-redshift power spectrum}, as follows: 

\begin{enumerate}  
\item If we keep fixed $h$, but now fix $\omcb$ instead of $\omm$, 
    as shown by the long-dashed line in Figure~\ref{fig-matpow},   
  then the small-scale power suppression is 
   somewhat reduced to 
  $\delta P / P \approx -6 \, \delta \xnu$, and we see a slight large-scale
  suppression $\delta P / P \approx -1 \, \delta \xnu$.  
 
\item If we consider (for simplicity) a pair of models with identical 
 $\omc, \omb, \xnu$ and identical early-time matter 
   power spectra in physical Mpc units, 
  but slightly different values of $h$, 
  the observables from a low-redshift galaxy survey are $P$ in units
  of $h^{-3} \Mpc^3$, and $k$ in units of $h \Mpc^{-1}$. 
 Defining $q = k / h$ and $\hat{P} = P/h^{-3}$, 
 the power-spectrum actually observed corresponds 
   to $\hat{P}(q) = h^3 P(k = hq)$ in units of $\Mpc^{-1}$ and Mpc$^3$.  
 Then, comparing two models 
  with identical $P(k)$ but a small difference in $h$, 
  two observers measuring the low-$z$ power spectrum 
  in these models would observe an offset in $\hat{P}(q)$ given by  
\begin{equation} 
  \delta \ln \hat{P}(q) \simeq
  \left[ 3 + \left( \frac{d \ln P}{d \ln k} \right)_{k = hq} \right] 
  \, {\delta \ln h}\ . 
\end{equation}  
 Given the $\Lambda$CDM power spectrum shape, the
  square-bracket term above is almost $+4$ on very large 
  scales ($k \simlt k_{eq}$), 
  then declines to $\sim +1$ at 
  $q \sim 0.5 \Mpc^{-1}$,  then asymptotes
  to zero on very small scales (where linear theory breaks down). 

 Since we saw above that the {\em Planck}-only degeneracy track 
  is well approximated 
 by $\delta \ln h \approx -2 \,\delta \xnu$,  the resulting 
  offset in $h$ contributes a fractional power spectrum shift 
\footnote{Effectively this is equivalent to a shift in the key large-scale
   structure parameter $\Omega_{cb}h$ \citep{esm90}: 
   if $\omcb$ is fixed,
  reducing $h$ produces an upward shift in $\Omega_{cb} h$, which moves 
  the power-spectrum rightwards in $q$.}  
 which is $\delta \ln \hat{P} \approx 
    -8 \, \delta \xnu$ at {\em small} $q$, 
  and ramps smoothly to $\sim -2$ at small scales 
   $q \sim 0.5 \, {\rm Mpc}^{-1}$. 
   By a rather remarkable apparent coincidence, this ramp from the $h-$offset 
  has similar magnitude but the opposite slope to the direct neutrino-mass 
   effect above: and the respective crossover scales 
   $k_{fs}$ and $k_{eq}$ also have a different origin but are 
   roughly similar for interesting neutrino masses; 
   so effects (i) and (ii) above 
   combine to produce a roughly uniform power 
   suppression $\delta \ln \hat{P}(q)  
   \sim -8 \, \delta \xnu$, now nearly independent of scale $q$. 

\item Finally, there is another shift in low-redshift power  
  due to the differing growth factor from the CMB era to today,  
   which depends mainly on $\Omega_m$:  
  the linear-theory $z = 0$ power spectrum contains a factor of $g^2$ where
   $g \propto \Omega_m^{0.24}$ is the linear-theory growth function. 
   Since we saw above that the CMB degeneracy track follows 
   $\delta \ln\Omm \approx  +5 \, \delta \xnu$, this 
  effect contributes a fractional shift in low-redshift $\hat{P}(q)$ by  
  $\delta \ln g^2 \approx 0.48 \, \delta \ln \Omm \approx 
   +2.4 \, \delta \xnu$, contributing an increase in power with $\xnu$; 
  this is in the opposite direction but smaller than effects 
  (i) and (ii) above.  
\end{enumerate} 
 
 { \newtwo (For a slightly different but comparable effect based on 
  super-sample density fluctuations, see \cite{li14}).}   

 Thus, the total effect of varying neutrino mass on the low-redshift 
  matter power spectrum is substantially dependent 
 on which other parameter(s) are held fixed:
  in Figure~\ref{fig-matpow} we show linear-theory 
  power spectra for three example cases, all
  with a common value $\delta \xnu = +0.01$ but different choices 
   for fixing other parameters.  

 To a fairly good approximation for $\xnu \simlt 0.03$, 
  since $\omcb$ and $\omb$ are almost unchanged we expect the high-redshift 
   power spectra in physical $k$ units 
  to vary only with effect (i) above; but at low
  redshift, effects (ii) and (iii) also contribute. In the approximation 
  that effects (i) - (iii) combine additively in $\ln \hat{P}$,  
   simply adding them predicts that varying neutrino mass (along the CMB 
  degeneracy track) results in an approximately scale-independent 
  suppression of the broad-band low-redshift power spectrum
  by a fractional shift  $\delta \ln \hat{P}(q) \sim -5.6 \, \delta \xnu$.  
 This approximate estimate is similar to the solid line in 
   Figure~\ref{fig-matpow}, except for the wiggles.   
 Since the key parameter $\sigma_8$ is measured in an $8 \hmpc$
  sphere, this also depends on $\hat{P}(q)$ and  
  thus $\delta \ln \sigma_8 \sim -2.8 \, \delta \xnu$.  
 This broadband overall power offset is largely degenerate 
  with the galaxy bias parameter in galaxy power spectra measurements,
   but the $\sigma_8$ effect is distinctive.  

 However, the BAO peaks do shift: along the {\em Planck} degeneracy track 
  the values of $\omcb$ and $\omb$ are almost independent of $\xnu$, so the BAO
 scale is almost independent of neutrino mass in physical $k$ units,  
  but it does shift in $q$ units due to the consequential
  shift in $h$; this shift is responsible for the pronounced wiggles in 
 the solid line in Figure~\ref{fig-matpow}.  

 Finally, we note another feature derived from the above: 
  weak-lensing measurements at moderate redshift are 
  especially sensitive to the parameter combination $S_8$, usually defined as 
   $S_8 \equiv \sigma_8 (\Omm/0.30)^{0.5}$.  
 Along the CMB degeneracy direction, 
   the combination of the $\sigma_8$ reduction as above with 
 the positive degeneracy $\delta \ln \Omm \sim +5 \, \delta \xnu$ 
  results in a near cancellation of the two effects on $S_8$;  
   again this is largely coincidental.  
  This helps to explain why the CMB constraints on $S_8$ are 
  counter-intuitively rather insensitive to varying neutrino mass
  (e.g. \citealt{maccrann15}).  

 To summarise this section, we have seen that when varying 
  neutrino mass along with other parameters 
   following the CMB degeneracy track, the ``secondary'' 
  effects on the matter power spectrum at large scales 
   caused by the consequential shifts in $\omlam$, $h$ and $\Omm$ are (mainly
  coincidentally) of a similar magnitude to the ``primary'' effect 
   of neutrinos suppressing small-scale power. 
 This explains qualitatively why BAO measurements and also $\sigma_8$ 
  measurements are in practice considerably more 
  effective than broad-band galaxy power spectrum 
 measurements for breaking the CMB-only neutrino mass degeneracy 
 (see e.g. \cite{cuesta16}).

\section{Eight-parameter models} 
\label{sec-8par} 

 The estimates in previous Sections assumed the six+one parameter
 flat $\Lambda$CDM + $\mnu$ model.  However, it is interesting to consider
  the effect of allowing an eighth free parameter such as dark energy
  equation of state $w$, curvature $\Omk$ or additional relativistic  
  species $(\neff > 3.046)$, since an extra free parameter may
   generally relax the upper limits on $\sumnu$. 
  Here we give just a short qualitative discussion of these three
   possible extra parameters in turn.   

 In the case of allowing $w \ne -1$, 
   it is well known that CMB fits give an anticorrelation 
  between $w$ and $H_0$ \citep{wein13}:
  assuming flatness,  increasing $w > -1$ requires lower $h$ 
 and higher $\Omega_m$ to fit the CMB data, i.e. the same direction 
  as increasing neutrino mass. Adding BAO measurements 
   gives primarily a constraint on $\Omcb$, hence implying an
  anticorrelation between neutrino mass and $w$ in a combined CMB+BAO fit.  
  This suggests that allowing ``phantom'' dark energy with $w < -1$ can relax 
   upper limits on neutrino mass, but allowing
  time-variable dark energy with a choice of a ``no-phantom'' prior 
  $w(z) \ge -1$, in general
  should not much weaken the upper limits on neutrino mass 
  (though if future CMB+BAO fits show a deviation from 6-parameter 
   $\Lambda$CDM, 
  there may well be potential ambiguity between the cases $\sumnu > 0.06 \ev$ 
   or $w > -1$). 
  
 For the case of small non-zero curvature $\Omk \ne 0$, 
  the largest change in observables 
   is a significant shift in the sound horizon angle, 
  with a high sensitivity 
  $\partial \ln \theta_* / \partial \Omega_k \approx -1.6$. 
  Since we have seen above that the sound horizon angle is a key factor 
  giving rise to the neutrino mass/dark energy degeneracy, 
  we expect that allowing non-flat models will significantly weaken
  the current constraints on neutrino mass, compared to assuming flatness. 
  This is consistent with the results of \cite{chen16}.  

 In the case of allowing $\neff > 3.046$,  we recall the argument
 of \cite{ew04} and \cite{suth12}: allowing free $\neff$  
   leads to a degeneracy direction whereby 
  to minimise changes in (dimensionless) CMB+BAO+SNe observables,  
  the physical densities of matter and vacuum energy increase 
   almost pro-rata with early-time radiation density. 
  Along this degeneracy track, 
   truly dimensionless parameters such as $\Omega_{cb}, \theta_*, z_{eq}$
  (which depend on density {\em ratios}) have best-fit values
   almost independent of $\neff$,  
   while the pseudo-dimensionless parameters $h$ and $\omcb$ 
  (which include an arbitrary normalisation to $H_0 = 100 \hunit$) 
    do show a substantial degeneracy with $\neff$.   
  Since neutrino mass has a substantial degeneracy with $\Omcb$
  but $\neff$ has little degeneracy with $\Omcb$,  this suggests 
    that allowing non-standard 
  $\neff$ will not substantially weaken neutrino mass limits from 
  dimensionless data combinations such as CMB+BAO+SNe; this is 
  broadly consistent with the results of \cite{planck15pars}. 

 {\newtwo As a numerical check, we have repeated the procedure from 
  Section~\ref{sec-num} of fitting a quadratic to 
    $\omlam$ vs $\omnu$ in a {\em Planck} MCMC
  chain, this time to the 8-parameter chain with 
  variable $\neff$ and $\mnu$ and selecting the subset of the chain
   with $3.4 < \neff < 3.6$, near the {\em Planck} upper limit.  
  This fit gave a slope of -14.7, only slightly steeper than the -14.4 
   found previously for standard $\neff$. }  
  
 To summarise this section, we estimate qualitatively that allowing 
  phantom dark energy ($w < -1$) 
  or non-zero curvature can substantially weaken the
  constraints on neutrino mass compared to the 7-parameter $\Lambda$CDM 
   + $\sumnu$ case; 
  but allowing $w > -1$ or non-standard $\neff$ 
  will tend to give only marginal weakening of neutrino mass upper limits
  from combined CMB+BAO+SNe datasets.   

\section{Conclusions} 
\label{sec-conc} 

We have given a simple and intuitive semi-analytic explanation 
 for the observed CMB degeneracy direction 
  in flat $\Lambda$CDM models extended with non-minimal neutrino mass. 
  A notable feature is that the key sound horizon angle is 
   about $2.3 \times$ {\em more} sensitive to small changes in 
   neutrino density than equivalent changes in 
  CDM density; this helps to explain  
  why the effects of small neutrino masses on the CMB cannot in practice 
  be compensated by the ``intuitive'' route of tweaking the CDM density.   
  Instead, to compensate the effect of increasing neutrino density 
   on the CMB requires a much larger reduction in vacuum energy, 
   $\delta \omlam \approx -14 \, \delta \omnu$, and
  this propagates into many other parameters:   
  we derived approximate dimensionless multipliers relating shifts  
 in the neutrino/matter ratio $\xnu \equiv  \omnu / \omcb$ 
  to consequential shifts in the ``low-redshift'' cosmological parameter(s)  
 ($H_0, \Omm, \Omlam$) along the CMB degeneracy track.   
These multipliers can be straightforwardly understood 
  from approximation~(\ref{eq:dstar3}) 
 and show good agreement with the numerical likelihood results from the
  {\em Planck} team.  

 A notable point is that a non-cosmological estimate of vacuum energy 
  density $\omlam$,  either from a future laboratory detection 
  or an {\em ab initio}  theoretical calculation, 
  could give strong constraints on 
  neutrino mass; unfortunately at present there is no well-agreed route to 
 such an estimate, though there are some speculative 
   proposals (e.g. \citealt{hogan12}; \citealt{padmanabhan16}). 
 
 We also gave an approximate explanation how the degeneracy between
  $\sumnu$ and $h$  produces a suppression in large-scale power in 
   observable $h$-dependent units; 
  when combined with the small-scale effect of neutrino mass, this
   explains the nearly scale-independent suppression of broadband 
  power at low redshift, combined with a sideways shift in the BAO 
  features.   

% Qualitatively, since $\omcb$ is constrained
%  very well by the CMB, increasing $\omnu$ requires a dramatically
% larger reduction in physical vacuum energy $\omlam$ in 
%  order to conserve the CMB acoustic angle $\theta_*$,
%  and this reduction propagates into all the other low-redshift parameters 
%  e.g. $\Omm$, $h$.  
  These multipliers above are helpful to intuitively explain 
  the strong constraints on total neutrino mass obtained from adding 
  low-redshift cosmological observations such as BAOs
  to the {\em Planck} data.  

\section*{Acknowledgements}

{\newtwo We thank the anonymous referee for helpful comments which significantly
 improved the clarity of the paper}.  We acknowledge the use of data from 
 the Planck Legacy Archive 
  supported by ESA at {\tt planck.esa.int}. 

% \newpage 
 
%% RRRRRRRRRRRRR tag for refs. 
%% now updated with MNRAS standard \abbrevs. 

%% AAAAAA tag for appendix. (no appendix). 
%\onecolumn 
%\appendix

\end{document}